\documentclass[prl,twocolumn,showpacs,amsmath,amssymb,citeautoscript,bibnotes,preprintnumbers]{revtex4-1}

\usepackage{graphicx}
\usepackage{bm}
\usepackage{amsfonts}

\newcommand{\psib}{{\overline{\psi}}}

\newcommand{\beq}{\begin{equation}}
\newcommand{\eeq}{\end{equation}}
\newcommand{\beqa}{\begin{eqnarray}}
\newcommand{\eeqa}{\end{eqnarray}}

\begin{document}

\title{Fermion bags, duality and the three dimensional massless lattice Thirring model}
\author{Shailesh Chandrasekharan} 
%\email{sch@phy.duke.edu}
\affiliation{Department of Physics, Box 90305, Duke University,
Durham, North Carolina 27708, USA}
\author{Anyi Li}
%\email{anyili@uw.edu}
\affiliation{Institute for Nuclear Theory, University of Washington, Seattle, WA 98195-1550, USA}
\preprint{INT-PUB-11-059}

\begin{abstract}
The recently proposed fermion bag approach is a powerful technique to solve some four-fermion lattice field theories. Due to the existence of a duality between strong and weak couplings, the approach leads to efficient Monte Carlo algorithms in both these limits. The new method allows us for the first time to accurately compute quantities close to the quantum critical point in the three dimensional lattice Thirring model with massless fermions on large lattices. The critical exponents at the quantum critical point are found to be $\nu=0.85(1)$, $\eta = 0.65(1)$ and $\eta_\psi = 0.37(1)$.
\end{abstract}


\maketitle

Strongly interacting quantum critical points containing massless fermionic excitations are interesting from many perspectives. Such critical points describe second order phase transitions in  strongly correlated two dimensional materials such as Graphene \cite{herbut:146401}. The search for critical points in gauge theories in three spatial dimensions containing massless fermions is interesting in the context of the physics beyond the standard model~\cite{Appelquist:2007hu}. Properties of these fermionic quantum critical points remain poorly understood due to lack of reliable computational techniques. While the renormalization group approach based on large $N$ or $\varepsilon$-expansion techniques can provide a glimpse of the rich possibilities, accurate Monte Carlo (MC) calculations are desirable. Unfortunately, MC methods for lattice field theories with massless fermions in three or more space-time dimensions continue to be challenging. For example, the popular Hybrid Monte Carlo (HMC) method \cite{PhysRevB.36.8632} either suffers from sign problems or encounters severe  singularities due to small eigenvalues of the fermion matrix. All HMC calculations performed so far have always relied on extrapolations to the massless limit which are known to be difficult~\cite{Sharpe:2007yd}. As far as we know MC calculations close to a quantum critical point with exactly massless fermions on large lattices do not exist.

Recently, a new approach called the fermion bag approach was proposed as an alternative method to solve a class of lattice field theories with exactly massless fermions~\cite{Chandrasekharan:2009wc}. It is an extension of the meron cluster idea proposed some time ago \cite{PhysRevLett.83.3116}. The idea behind the fermion bag is to identify fermion degrees of freedom that cause sign problems and collect them in a bag and sum only over them. This is in contrast to traditional approaches where all fermion degrees of freedom in the entire thermodynamic volume are summed to solve the sign problem. When the fermion bag contains only a small fraction of all the degrees of freedom and the summation can be performed quickly, the fermion bag approach can be used to design powerful MC methods. Sometimes, the bag splits into many disconnected pieces further simplifying the calculation.

The general idea of the fermion bag can be illustrated easily with the following example. Consider lattice fermion models formulated with $2n$ Grassmann variables per lattice site denoted as $\psi_i(x)$ and $\psib_i(x)$, where $i=1,2,..n$ represent flavor indices and $x$ denotes the Euclidean space-time lattice point containing $V$ sites. Let $D$ be the $V \times V$ free fermion matrix whose matrix elements are denoted as $D_{xy}$. We will assume that the properties of $D$ are such that the following $k$-point correlation function involving the flavor $i$ :
\begin{eqnarray}
C_i(x_1,...,x_k) &=& \int [d\psib d\psi] \exp\Big(\sum_{x,y} \ \psib_i(x) \ D_{xy}\ \psi_i(y) \Big)\nonumber \\
&& \ \ \  \psib_{i}(x_1)\psi_{i}(x_1)\ ...\ \psib_{i}(x_k)\psi_{i}(x_k)
\end{eqnarray}
is always positive. An example of such a matrix $D$, is the massless staggered fermion Dirac operator which is popular in constructing four-fermion lattice models \cite{AnnPhys.224.29}. It is easy to prove that
\begin{equation}
C_i(x_1,..,x_k) = \mathrm{Det}(D)\ \ \mathrm{Det}(G[\{x\}_i])
\label{smallu}
\end{equation}
where $G[\{x\}_i]$ is the $k \times k$ matrix of propagators between the $k$ sites $x_p,p=1,..,k$ whose matrix elements are $G_{x_p,x_q} = D^{-1}_{x_p,x_q}$. It is also possible to argue that \cite{Chandrasekharan:2009wc},
\begin{equation}
C_i(x_1,..,x_k) = \mathrm{Det}(W[\{x\}_i])
\label{largeu}
\end{equation}
where the matrix $W[\{x\}_i]$ is a $(V-k) \times (V-k)$ matrix which is the same as the matrix $D$ except that the sites $\{x\} \equiv \{x_p,p=1,2,..k\}$ are dropped from the matrix. The identity
\begin{equation}
\mathrm{Det}(D)\ \ \mathrm{Det}(G[\{x\}_i]) \ =\ \mathrm{Det}(W[\{x\}_i])
\label{duality}
\end{equation}
leads to a concept of duality in the fermion bag approach as we explain below.

\begin{figure*}[t!]
\begin{center}
\hbox
{\hskip 0.8in
\includegraphics[width=0.37\textwidth]{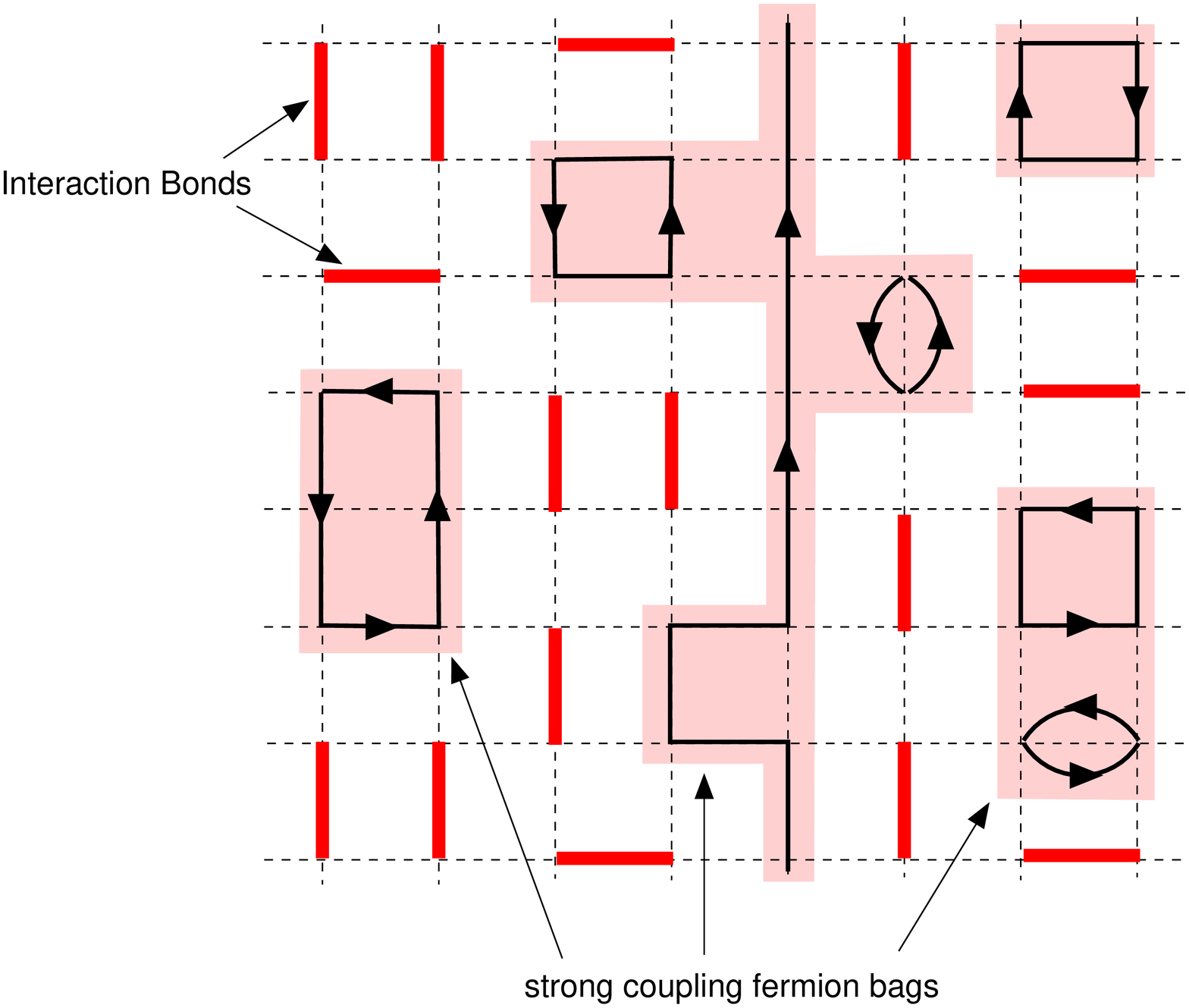}
\hskip .5in
\includegraphics[width=0.34\textwidth]{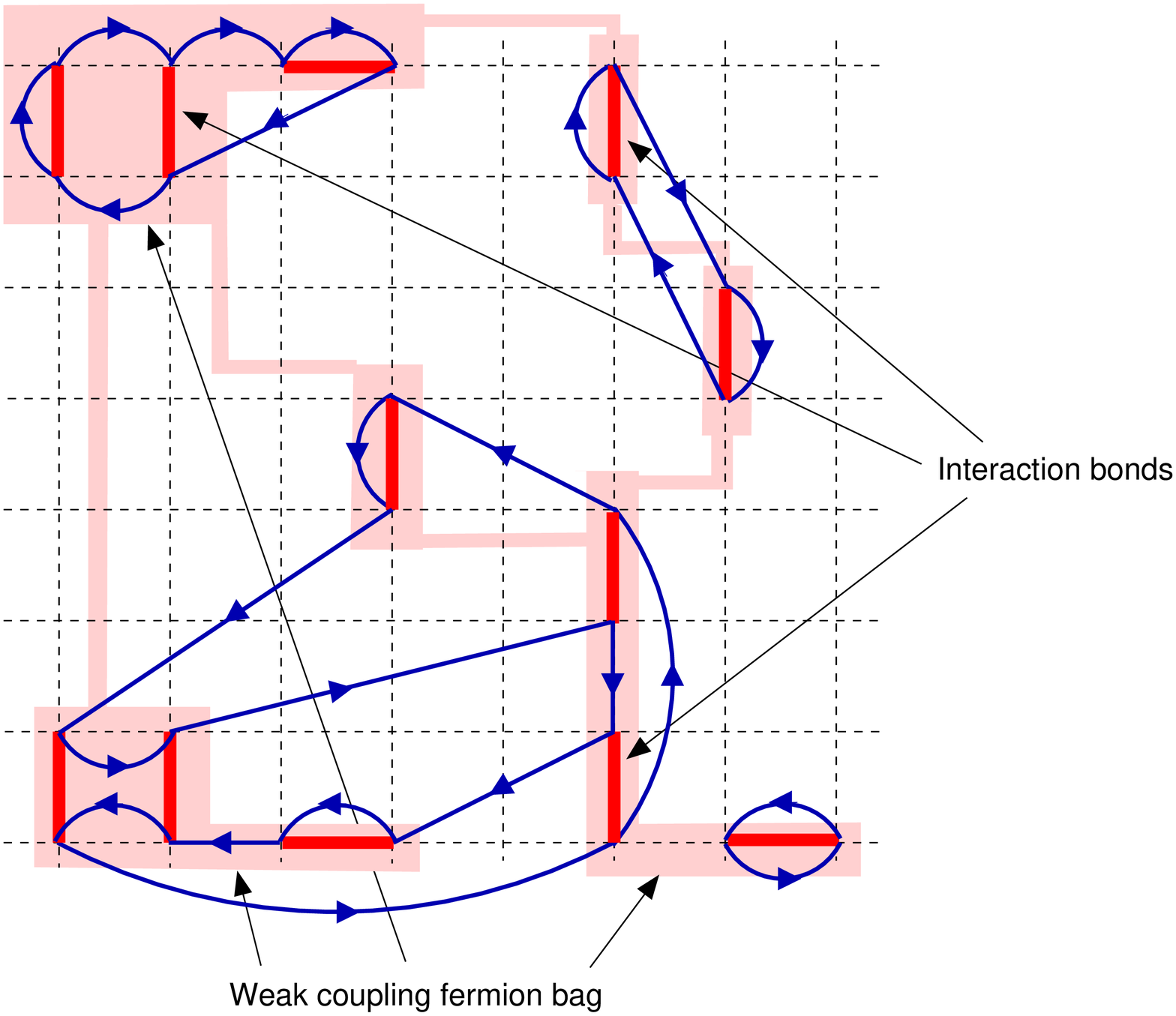}
}
\end{center}
\caption{\label{fig:bag} An illustration of a ``fermion-bag'' configuration at strong couplings (left) and weak couplings (right). The interactions in this illustration are represented by solid bonds and the fermion bags are represented by shaded region. At strong couplings the fermion bag is made up of free sites and breaks up into many disconnected pieces, while at weak couplings the bag contains interaction sites.}
\end{figure*}

Consider a generic four-fermion lattice field theory action involving $n$ massless staggered fermions given by
\begin{eqnarray}
S &=& - \sum_{x,y,i } \ \psib_i(x) \ D_{xy} \psi_i (y) \nonumber \\
&& - \sum_{\langle xy \rangle} \sum_{i,j} \ U_{i,j,\langle xy \rangle} \psib_i(x)\psi_i(x)  \psib_j(y)\psi_j(y)  
\end{eqnarray}
where $\langle xy\rangle$ refers to some well defined set of neighboring lattice sites. Further we will assume that all the couplings $U_{i,j,\langle xy\rangle }$ are non-negative real constants. Many interesting four-fermion models are of this type \cite{AnnPhys.224.29}. The partition function can be expanded in powers of the coupling and is given by
\begin{equation}
Z = \int \ [d\psib d\psi] \  \mathrm{e}^{ -S} = \sum_{[\{x\}]} \{[U]^p\}\ \prod_{i=1}^n C_i(x_1,...,x_{k_i}),
\end{equation}
where $\{[U]\}^p$ refers to a generic power of the coupling and $k_i,i=1,2,..n$ refers to the number of interaction vertices for each flavor. On a finite lattice the expansion is convergent since it is a polynomial.

The above expansion of the partition function begs for the following intuitive interpretation. Since the $k_i$ interaction sites contain both $\psi_i$ and $\psib_i$, the $i^{\rm th}$ flavor of fermions are already paired on these sites and do not cause sign problems. On the other hand, unpaired fermions of the $i^{\rm th}$ flavor that hop freely on the remaining sites can indeed cause sign problems and need to be summed over to solve the sign problem. The free sites are collectively referred to as a fermion bag. The summation of all fermion world lines inside the bag leads to the weight  $C_i(x_1,x_2,..,x_k) = \mathrm{Det}(W[\{x\}_i])$ which is the determinant of a $(V-k_i) \times (V-k_i)$ matrix. This determinant can be evaluated easily if $(V-k_i)$ is small. This is expected at strong couplings and hence we refer to these free fermion bags as strong coupling fermion bags. It was shown in \cite{Chandrasekharan:2009wc} that at strong couplings a fermion bag splits into many small disconnected pieces making things even simpler. The left figure of Fig.~\ref{fig:bag} gives an illustration of the disconnected pieces of a strong coupling fermion bag.

At weak couplings the above definition of a fermion bag loses its charm since $V-k_i$ becomes large. However, thanks to a concept of duality we can construct the fermion bag differently. At weak couplings we can view the interactions as the unpaired fermionic degrees of freedom that cause fluctuations over the paired free fermionic vacuum. In this view the fermions hop from one interaction site to another interaction site leading to sign problems and need to be summed over. Now the interaction sites form the fermion bag. Again the summation of the fermions inside this dual bag leads to the same weight $C_i(x_1,x_2,..,x_k) = \mathrm{Det}(W[\{x\}_i]) = \mathrm{Det}(D)\ \ \mathrm{Det}(G[\{x\}_i])$ but now viewed as the determinant of a $k_i \times k_i$ matrix. Note we have used the duality relation, Eq.(\ref{duality}) here. The determinant can now be calculated easily since $k_i$ is small at weak couplings. Hence we refer to these dual bags as weak coupling fermion bags. The right figure of Fig.~\ref{fig:bag} gives an illustration of a weak coupling fermion bag. The weak coupling fermion bag approach is equivalent to the idea of summing over all Feynman diagrams and was introduced earlier in the framework of diagrammatic determinantal Monte Carlo method ~\cite{RevModPhys.83.349}. On the other hand, in our opinion the fermion bag approach is more intuitively appealing in the context of lattice field theories since it uncovers the powerful concept of duality and extends to strong couplings.

The fermion bag approach is general and can be adapted to relativistic Wilson fermions and non-relativistic fermions. However, in some models the weight of a fermion bag is no longer a determinant, but involves new mathematical structures like fermionants~\cite{Chandrasekharan:2011an}, which can be exponentially difficult to compute \cite{MertensS2011}. In such cases the fermion bag approach is not useful. However, these models can be changed by introducing unconventional interactions (like six or higher fermion interactions), while still preserving many interesting symmetries and the corresponding low energy physics \cite{Chandrasekharan:2009wc}. In these exotic models, the fermion bag weight is again a determinant and the fermion bag approach becomes useful. Thus, many new fermion models can be solved with the fermion bag approach.

As a first application of the new approach we have studied the three dimensional massless lattice Thirring model with two Grassmann valued fields per site denoted by $\psi(x)$ and $\psib(x)$. The action is given by
\begin{equation}
S = -\sum_{x,y} \ \psib(x) \ D_{xy}\ \psi(y) - U \sum_{\langle x y\rangle}  \ 
\psib(x) \psi(x) \ \psib(y)\psi(y),
\label{eq:model}
\end{equation}
where $D$ is the massless staggered fermion matrix, $\langle xy\rangle$ refers to the set of nearest neighbor sites of a cubic lattice and $U$ is the coupling that generates the current-current coupling of the continuum Thirring model. We use anti-periodic boundary conditions in all three directions. The lattice model is invariant under a $U_f(1)\times U_\chi(1)$ symmetry, where $U_f(1)$ is the fermion number symmetry and $U_\chi(1)$ is the chiral symmetry. When the coupling is small the model contains four flavors of massless two-component Dirac fermions at long distances due to fermion doubling. When the coupling is large, chiral symmetry breaks spontaneously and generates a single massless Goldstone boson and the fermions become massive. Thus, the model contains a quantum critical point $U_c$ which separates a phase with massless fermions from a phase with massless bosons. The quantum critical point has been studied earlier using mean field techniques~\cite{PhysRevLett.59.14}, and traditional MC methods~\cite{DelDebbio:1995zc,DelDebbio:1997dv,DelDebbio:1997hd,Barbour:1998yc}. A variant of our lattice model has been used recently to study a related quantum critical point in the context of Graphene~\cite{PhysRevB.84.075123}.

Close to the quantum critical point a continuum quantum field theory description of the long distance physics must emerge. This continuum theory should contain four flavors of two component Dirac fermions in three Euclidean dimensions. As was discussed in \cite{DelDebbio:1997dv}, the lattice interactions generate many continuum four-fermion interaction terms and the continuum Lagrange density takes the form
\begin{eqnarray}
&& {\cal L} \ =\ \psib_i(x) (\vec{\sigma} \cdot \vec{\nabla}) \psi_i(x) 
+ \  \Big\{g_A\big[\psib_i(x) \vec{\sigma}  (\Gamma_A)_{ij} \psi_j (x)\big]^2 
\nonumber \\
&& 
\ \ \hskip1in \ +\  \tilde{g}_A\big[\psib_i(x) (\Gamma_A)_{ij}\psi_j(x)\big]^2\Big\}
\end{eqnarray}
where $\psib_i(x), \psi_i(x),i=1,..,4$ are the four flavors of two component Dirac fermion fields, $\vec{\sigma}$ are the three Pauli matrices, $\Gamma_A, A=1,..,16$ are the sixteen generators of the $U(4)$ group in the flavor space under which the free theory is invariant. Repeated indices are assumed to be summed over. The couplings $g_A$ and $\tilde{g}_A$ are only constrained by the lattice symmetries. They take values such that the $U(4)$ symmetry of the free theory is broken to a $U_f(1) \times U_\chi(1)$ subgroup \footnote{Some additional discrete symmetries may also be preserved on the lattice.}. As far as we know, a renormalization group (RG) flow analysis in this relatively large but constrained space of couplings, within a controlled approximation such as large $N$ or $\varepsilon$-expansion, is not available and should be an interesting topic for future research. However, the existence of a quantum critical point in the lattice model does imply that the RG analysis will find a nontrivial fixed point with one relevant direction. Here we compute the critical exponents at this fixed point through MC calculations. 

The fermion bag approach for the above lattice model was first developed in~\cite{Chandrasekharan:2009wc} and it was shown that the partition function can be written as
\begin{equation}
Z = \sum_{[n]} U^{N_b} \mbox{Det}(W[n])
\end{equation}
where $[n]$ refers to the configuration of $N_b$ interaction bonds and $W[n]$ is the $(V-2N_b)\times (V-2N_b)$ staggered Dirac matrix restricted to the free sites. These free sites form the strong coupling fermion bag  (see left figure in Fig.~\ref{fig:bag}). At the quantum critical point $N_b$ is about an eighth of the lattice volume and hence the above form of the partition function is not useful. However, thanks to duality we can think in terms of the weak coupling fermion bags (see right figure of Fig.~\ref{fig:bag}). For the above model, the duality relation (Eq.~(\ref{duality})) takes the form
\begin{equation}
\mbox{Det}(W[n]) = \mbox{Det}(D)[\mbox{Det}^2(G[n])],
\end{equation}
where $G[n]$ is an $N_b \times N_b$ free fermion propagator matrix between even and odd lattice sites of the interaction bonds.  Using recent algorithmic advances \cite{Adams:2003cc}, we have constructed an efficient determinantal Monte Carlo algorithm for this problem \cite{Chandrasekharan:2011vy}. If the autocorrelation times and equilibration times are measured in units of a sweep, we find that our algorithm has no further critical slowing down even at the quantum critical point.

\begin{figure*}[t]
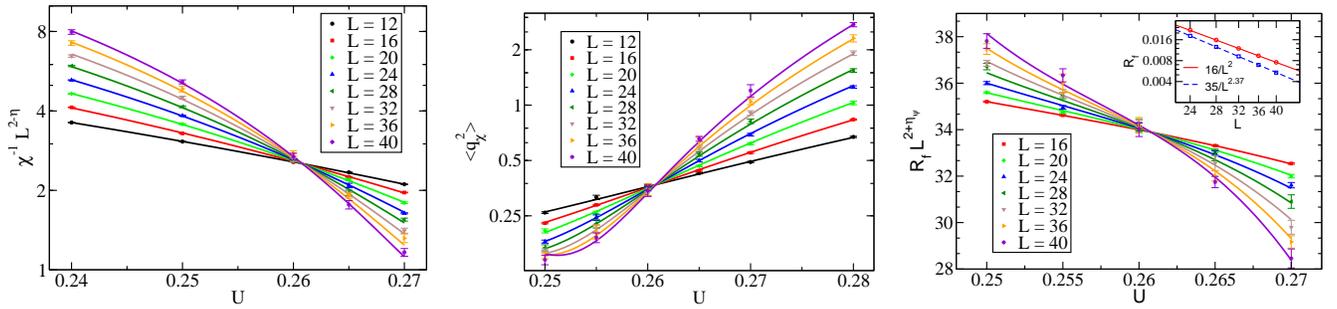

\begin{center}
\hbox{
\includegraphics[width=2.2in]{./figs/z1.eps}
\hskip0.1in
\includegraphics[width=2.2in]{./figs/susc.eps}
\hskip0.1in
\includegraphics[width=2.2in]{./figs/fratio.eps}
}
\caption{Plots of $\chi^{-1} L^{2-\eta}$, $\langle q_\chi^2\rangle$ and $R_f L^{2+\eta_\psi}$ as a function of $U$ for $L$ from $12$ to $40$. The solid lines show the combined fit which give $U_c = 0.2608(2), \nu = 0.85(1), \eta=0.65(1)$ and $\eta_\psi=0.37(1)$. The inset shows a linear-log plot of $R_f$ versus $L$ at $U=0$ (solid line) and $U=0.260$ (dashed line). The lines are fits to a power law. }
\label{fig:scaling}
\end{center}
\end{figure*}

In order to uncover the properties of the quantum critical point we focus on three observables (Let $L$ be the lattice size): The chiral condensate susceptibility \footnote{due to a technical difficulty, instead of $\chi$ we measure a related quantity. The difference is explained in \cite{Chandrasekharan:2011vy}},
\beq
\chi = \frac{1}{2 L^3}\sum_{x,y}\langle\psib_x\psi_x\psib_y\psi_y\rangle,
\eeq
the chiral winding number susceptibility 
\beq
\langle q^2_\chi \rangle =\langle \frac{1}{3}\sum_\alpha(q^2_\chi)_\alpha \rangle,
\eeq 
(defined through the conserved chiral charge $(q_\chi)_\alpha=\sum_{x \in S} \ \varepsilon_x\ \eta_{x,\alpha}\ (D^{-1})_{x,x+\alpha}\ +\ \sum_{x\in S}\  2\varepsilon_x$ passing through the surface $S$ perpendicular to the direction $\alpha$,  where  $\varepsilon_x = (−1)^{x_1+x_2+x_3}$. The staggered fermion phase factors $\eta_{x,\alpha}=\exp(i\pi\xi_a \cdot x)$ are defined through the 3-vectors $\xi_1=(0,0,0)$, $\xi_2=(1,0,0)$ and $\xi_3=(1,1,0)$), and the ratio 
\beq
R_f=C_F(L/2-1)/C_F(1), 
\eeq
(defined through the fermion two point function $C_F(d)=\frac{1}{3}\sum_{\alpha = 1}^3 \langle\bar{\psi}_x\psi_{x+d\hat{\alpha}}\rangle$ in which $x$ belongs to a site with $\varepsilon_x = 1$ and $\hat{\alpha}$ is a unit vector along each of the three directions). Since the fermions are exactly massless, in the vicinity of $U_c$ we expect these three observables to satisfy the following simple finite size scaling relations:
\begin{subequations}
\label{eq:scaling}
\beqa
\chi^{-1} L^{2-\eta} = \sum_{k=0}^3 f_k \left[(U-U_c) L^{\frac{1}{\nu}}\right]^k \\
\langle q_\chi^2 \rangle = \sum_{k=0}^3 \kappa_k \left[(U-U_c) L^{\frac{1}{\nu}}\right]^k \\
R_f L^{2+\eta_\psi} = \sum_{k=0}^3 p_k \left[(U-U_c) L^{\frac{1}{\nu}}\right]^k 
\eeqa
\end{subequations}
where we have kept the first four terms in the Taylor series of the corresponding analytic functions. Our goal is to compute the critical exponents $\eta$, $\nu$ and $\eta_\psi$ at the quantum critical point. Large $N$ calculations usually give $\nu = \eta = 1$ and $\eta_\psi = 0$ \cite{AnnPhys.224.29}.

Earlier studies of the quantum critical point were focused on computing $\nu$ and $\eta$. In these studies the four fermion coupling is converted into a fermion bilinear using an auxiliary field. The fermions are then integrated out and the remaining problem is solved using the HMC method. In order to avoid singularities, all calculations are done with a finite fermion mass. A close examination of the earlier work reveals that different analysis produce substantially different results. However, the presence of large errors makes everything look consistent. 
%For example one of the earlier work finds $\nu = 0.71(4)$ and $\eta=0.60(2)$ \cite{DelDebbio:1997hd} while another finds $\nu=0.88(8)$ and $\eta = 0.46(11)$ \cite{Barbour:1998yc}. The former result is obtained from calculations up to $16^3$ lattices, while the latter uses lattices up to $24^3$.  
In our opinion, the presence of two infrared scales (the fermion mass and the length of the box) makes the analysis difficult. In contrast, since we work with massless fermions, a single combined fit to Eqs.~(\ref{eq:scaling}) with sixteen parameters is easy.  Indeed, a combined fit of all our data from $12^3$ to $40^3$ lattice gives us $\nu = 0.85(1), \eta=0.65(1)$, $\eta_\psi=0.37(1)$ and $U_c = 0.2608(2)$ with a  $\chi^2/d.o.f.=1.3$. The complete list of the sixteen fit parameters are listed in Table~\ref{table:fit}. Our computation of $\eta_\psi$ is new.

\begin{table}[h]
\begin{center}
\begin{tabular}{c c c c c c}
\hline\hline
$f_0$ & $f_1$ &$f_2$& $f_3$ & $\kappa_0$ & $\kappa_1$ \\
\hline 
2.52(3) & -2.53(5) & 0.71(3) & 0.10(1) &0.369(3) & 0.63(1)   \\
\hline\hline
 $\kappa_2$& $\kappa_3$ &  $p_0$ & $p_1$& $p_2$& $p_3$ \\
\hline
0.52(2) & 0.09(1) & 33.9(2) & -5.0(1) & -2.0(2) & -2.5(5)\\
\hline\hline
\end{tabular} 
\caption{Results from the combined fit of the data to Eqs.~(\ref{eq:scaling}). In addition to the above twelve parameters, the fit also gives $U_c=0.2608(2)$, $\eta=0.65(1)$, $\nu=0.85(1)$, $\eta_\psi = 0.37(1)$ with a $\chi^2$/d.o.f of $1.3$.}
\label{table:fit}
\end{center}
\end{table}

Plots of our data along with the fits are shown in Fig.~\ref{fig:scaling}.  Since our data fits very well to the expected scaling form for a whole range of lattice sizes, we feel confident that the corrections to scaling are small. 

We thank Ph. deForcrand, S.~Hands, D.~Kaplan, K.-F.~Liu, C.~Strouthos and U.-J.~Wiese for discussions. This work was supported in part by the DoE grants DE-FG02-05ER41368 and DE-FG02-00ER41132.

\bibliography{ref}

\end{document}